\documentclass[twocolumn,showpacs,superscriptaddress,amsmath,amssymb]{revtex4}

\usepackage{graphicx}
\begin{document}

\title{Effects of electron-phonon coupling in angle-resolved photoemission spectra of SrTiO$_3$}

\author{K.~Maekawa}
\affiliation{Department of Physics, University of Tokyo, 
Bunkyo-ku, Tokyo 113-0033, Japan}
\author{M.~Takizawa}
\affiliation{Department of Physics, University of Tokyo, 
Bunkyo-ku, Tokyo 113-0033, Japan}
\author{H.~Wadati}
\affiliation{Department of Physics, University of Tokyo, 
Bunkyo-ku, Tokyo 113-0033, Japan}
\author{T.~Yoshida}
\affiliation{Department of Physics, University of Tokyo, 
Bunkyo-ku, Tokyo 113-0033, Japan}
\author{A.~Fujimori}
\affiliation{Department of Physics, University of Tokyo, 
Bunkyo-ku, Tokyo 113-0033, Japan}
\author{H.~Kumigashira}
\affiliation{Department of Applied Chemistry, University of Tokyo, 
Bunkyo-ku, Tokyo 113-8656, Japan}
\author{M.~Oshima}
\affiliation{Department of Applied Chemistry, University of Tokyo, 
Bunkyo-ku, Tokyo 113-8656, Japan}
\date{\today}

\begin{abstract}
We have studied the O $2p$ valence-band structure of Nb-doped SrTiO$_3$, in which a dilute concentration of electrons are doped into the $d^0$ band insulator, by angle-resolved photoemission spectroscopy (ARPES) measurements. 
We found that ARPES spectra at the valence band maxima at the M $\left[{\bf k} = \left(\frac{\pi}{a},\frac{\pi}{a},0\right) \right]$ and R $\left[{\bf k} = \left(\frac{\pi}{a},\frac{\pi}{a},\frac{\pi}{a}\right)\right]$ points start from $\sim$ 3.3 eV below the Fermi level ($E_F$), consistent with the indirect band gap of 3.3 eV and the $E_F$ position at the bottom of the conduction band. 
The peak position of the ARPES spectra were, however, shifted toward higher binding energies by $\sim 500$ meV from the 3.3 eV threshold. 
Because the bands at M and R have pure O $2p$ character, we attribute this $\sim 500$ meV shift to strong coupling of the oxygen $p$ hole with optical phonons in analogy with the peak shifts observed for $d$-electron photoemission spectra in various transition-metal oxides. 
\end{abstract}


\maketitle
\begin{figure*}
\begin{center}
\includegraphics[width=.9\linewidth]{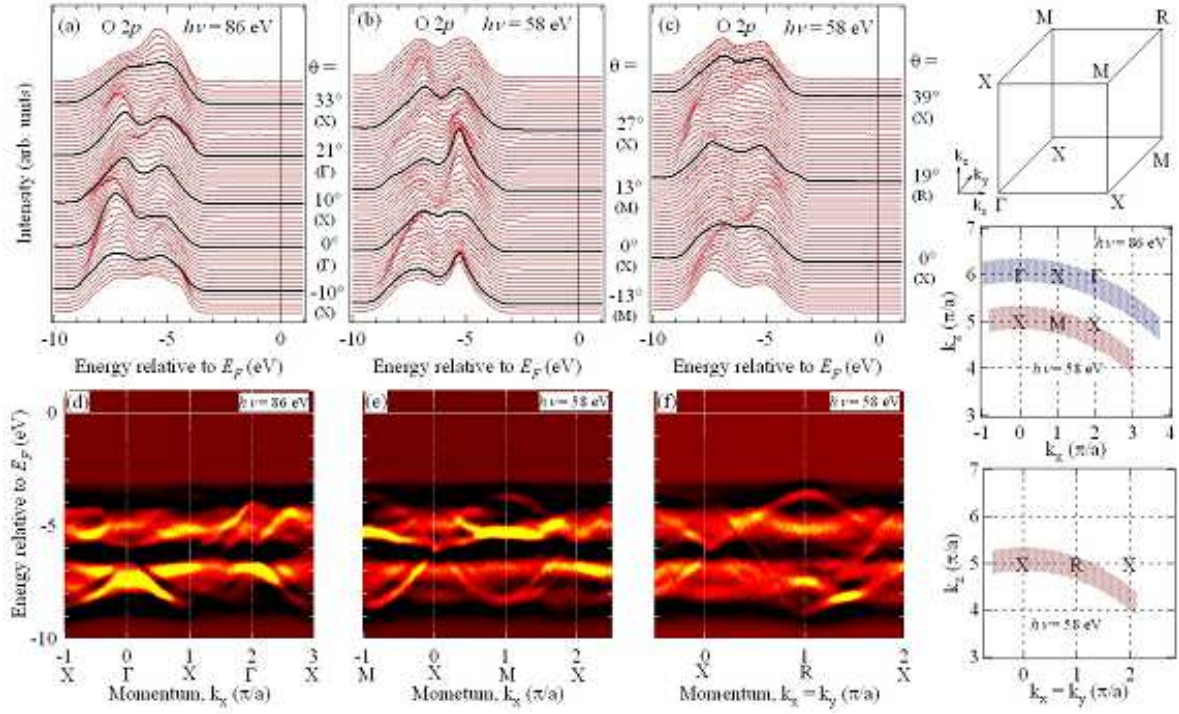}
\caption{(Color online) ARPES spectra of SrTiO$_3$. (a), (b), (c): EDCs taken with photon energies  $h\nu = 86 $eV and $h\nu = 58 $eV. 
(d), (e), (f): Gray-scale plots of the second derivatives of the EDCs for the $\Gamma$ - X direction, the X - M direction, and the X - R direction. 
Here, bright parts correspond to energy bands. 
Right top: Brillouin zone of the simple cubic lattice. 
Right middle and bottom: traces in momentum space for $h\nu$ = 86 eV and 58 eV and the emission angle from $\theta = -10^{\circ}$ to $40^{\circ}$. }
\label{STOgxEDC}
\end{center}
\end{figure*}

\begin{figure}[!b]
\begin{center}
\includegraphics[width=.9\linewidth]{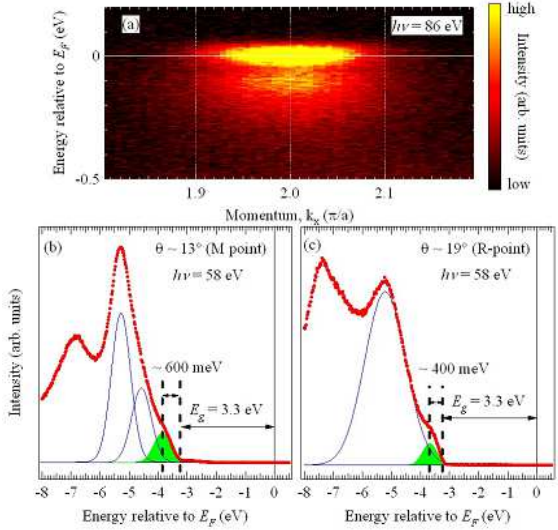}
\caption{(Color online) ARPES spectra of SrTiO$_3$. (a) Intensity plot in $E$-$k$ space around the $\Gamma$ point. The bottom of the Ti $3d$ band was observed at the $\Gamma$ point in the vicinity of $E_{F}$. 
(b), (c): EDCs at the M point and the R point. 
They have been decomposed into several Gaussians, out of which only the lowest binding energy ones are shown. }
\label{rm}
\end{center}
\end{figure}

Transition-metal oxides (TMO) are known to exhibit a variety of attractive phenomena such as high-temperature superconductivity, giant magnetoresistance and metal-insulator transition (MIT) \cite{MIT} resulting from strong interaction among the transition-metal $3d$ electrons. 
Photoemission spectroscopy (PES) is a powerful technique to study the electronic structure of such strongly correlated systems. 
Recently, it has become recognized that electron-phonon coupling also affects the PES spectra of the TMOs significantly. 
Recent PES measurements have demonstrated pronounced electron-phonon interaction effects in Cu oxides \cite{k.shen}, Fe oxides \cite{Wadati-LSFOarpes}, and V oxides \cite{okazaki}. 
Mishchenko and Nagaosa \cite{mishchenko} and R\"osch {\it et al}. \cite{Rosch} have theoretically shown that electron-phonon coupling affects the experimentally measured angle-resolved photoemission spectroscopy (ARPES) spectra of high-$T_c$ superconductors and their parent insulators through calculations based on the $t$-$J$ model with Holstein-type electron-phonon interaction. 
While those studies have focused on the effects of electron-phonon coupling primarily on transition-metal $d$ electrons, effects of electron-phonon coupling on oxygen $p$ electrons have not been studied so far. 
Since the $3d$ electrons and the oxygen $p$ electrons are strongly hybridized with each other in TMOs, it is of fundamental interest and importance to see how electron-phonon coupling affects the photoemission spectra of oxygen $p$ electrons. 

In the present work, we have performed an ARPES study of SrTiO$_3$ (STO), which is a perovskite-type $3d^0$ band insulator with the filled O $2p$ band and the empty Ti $3d$ band and is therefore the {\it starting point} of the series of the perovskite TMO’s. 
The band gap ({\it E$_{\mbox g}$}) of STO determined by optical measurements is $\sim 3.3$ eV for the indirect gap and $\sim 3.8$ eV for the direct gap \cite{Goldschmidt, vonBenthema}. 
Substituting La for Sr, Nb for Ti or introducing oxygen vacancies leads to electron doping and makes the system metallic and even superconducting already from very low carrier concentrations of $8.5 \times 10^{18}$ cm$^{-3}$ \cite{STO-SC1, STO-SC2}. 
Electron-phonon coupling in STO has long been studied. 
The optical and dielectric properties of STO have provided evidence for pronounced effects of polarons \cite{gervais, eagles, ang}. 
Recently, Bi $et$ $al.$ \cite{Bi} suggested from an infrared reflectivity study that small polarons exist and play an important role in Nb-doped STO.　
The band structure has been studied theoretically by many groups \cite{mattheissPRB, mattheissPR, vonBenthema, kahn}. 
ARPES measurements have also been performed \cite{haruyama, Aiura} to study the band structure of STO, however, these measurements were made for several $k$ points in the Brillouin zone, mainly using the so-called normal emission method. 
In the present work, we have made detailed ARPES measurements on Nb-doped, i.e., lightly electron-doped, STO and have examined the dispersions of the O $2p$ band and the bottom of the Ti $3d$ band. 
The dispersions of the O $2p$ band has been well explained by tight-binding (TB) band-structure calculation if the band gaps are adjusted to the optical gaps of 3.3 eV and 3.8 eV. 
Photoemission starts from $\sim 3.3$ eV below the Fermi level ($E_F$), consistent with the indirect optical gap of 3.3 eV. 
However, the ARPES peak positions of the O $2p$ bands are located $\sim 500$ meV below the valence band maximum (VBM). 
We shall discuss the origin of the $\sim$ 500 meV shift in the context of strong electron-phonon coupling. 

ARPES measurements were performed at BL-1C of Photon Factory, High Energy Accelerators Research Organization (KEK). 
A Nb-doped STO single crystal with an atomically flat (001) surface of TiO$_2$ termination was measured \cite{kawasaki}. 
To introduce carriers (electrons) into the sample, it was annealed under an ultra high vacuum ($\sim$ $10^{-9}$ Torr) at 1273 K for 2 hours, and then transferred to the photoemission chamber without exposing it to air \cite{horiba}. 
Measurements were performed under an ultrahigh vacuum of $\sim$ $10^{-10}$ Torr at 20 K using a Scienta SES-100 electron-energy analyzer. 
The total energy resolution including the monochrometer was set to $\sim 60$ meV and $\sim 150$ meV near $E_F$ and in the valence-band region, respectively. 
The $E_F$ position was determined by measuring gold spectra. 
Traces in momentum space for changing emission angle are shown in the right panels of Fig. \ref{STOgxEDC}, when the widths represent the uncertainties in the $k_{\mbox z}$ due to the finite escape depth of photoelectrons \cite{wadatitbc}. 
In order to determine the momentum perpendicular to the sample surface, we have assumed the work function of the sample $\phi = 4.5$ eV, the inner potential $V_0 = 10.5$  eV, and the lattice constant $a = 3.905$ \AA. 

\begin{figure*}
\begin{center}
\includegraphics[width=\linewidth]{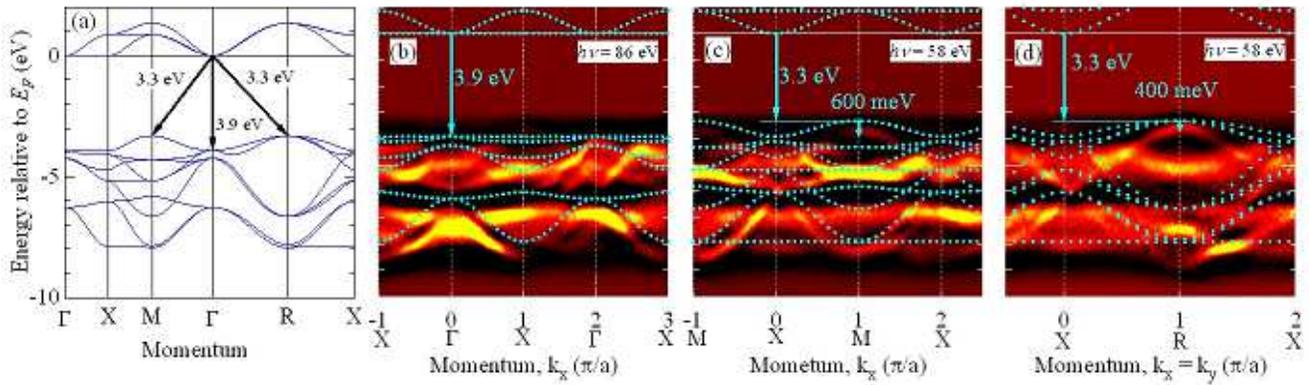}
\caption{(Color online) Comparison between the ARPES band structure and TB calculation, (a) Band structure of STO. 
(b), (c), (d): Comparison for the $\Gamma$ - X, the X - M, and the X - R directions. 
The direct and indirect gaps are indicated.
The TB calculation has reproduced the experimental ARPES band structure except for the overall shifts of the O 2$p$ band by $\sim$ 500 meV.}
\label{ARPESwithcalc}
\end{center}
\end{figure*}
Figure~\ref{STOgxEDC} (a), (b), and (c) shows energy distribution curves (EDCs) taken with photon energies of 86 eV and 58 eV, showing clear band dispersions. 
In order to see the band dispersions more clearly, we have taken the second derivatives of the EDCs and plotted them on a gray scale in Fig.~\ref{STOgxEDC} (d), (e), and (f). 
Here, bright part indicates peaks or shoulders in the EDC's. 
In Fig.~\ref{rm} (a), the $E$-$k$ intensity plot near $E_F$ around the $\Gamma$ point is shown. 
The intensity in the vicinity of $E_F$ arises from the bottom of the Ti $3d$ band, which is occupied by a small amount of electrons doped into the sample. 
From the size of the occupied part in momentum space ($|k_F| \sim 0.15 \frac{\pi}{a}$), the carrier density is estimated to be $\sim \times 10^{-2}$ per Ti atoms by assuming the Fermi surface to be three-fold degenerate spheres of the $t_{2g}$ bands. 
Then, the Fermi energy is estimated to be as small as $\sim 40$ meV from the above $|k_F|$ value and the effective mass $m^* = 1.5m_0$ \cite{haruyama}, where $m_0$ is the free-electron mass. 
Therefore, the Fermi level is located only $\sim 40$ meV above the bottom of the Ti $3d$-derived conduction band. 
The EDCs at the M point [${\bf k}=\left(\frac{\pi}{a},\frac{\pi}{a},0\right)$] and the R point [${\bf k}=\left(\frac{\pi}{a},\frac{\pi}{a},\frac{\pi}{a}\right)$], both corresponding to the top of the O $2p$ band of pure O $2p$ character, are shown in Fig.~\ref{rm} (b) and (c). 
The emission starts at the threshold of $\sim 3.3$ eV below $E_F$, that is, 3.3 eV below the bottom of Ti $3d$ band. 
The value of 3.3 eV is equal to the magnitude of the indirect band gap of STO between the valence-band maximum at the M or R point and the conduction-band minimum at the $\Gamma$ point. 
Both spectra have been decomposed into several Gaussians, out of which the lowest binding energy ones are shown in Fig.~\ref{rm} (b) and (c). 
Thus, the peak position of the lowest binding energy structure is observed 3.8 eV below $E_F$, i.e., by $\sim 500$ meV deeper than the threshold. 

In order to interpret the experimental band dispersions quantitatively, the ARPES spectra have been fitted to TB band-structure calculation. 
We have performed the TB calculation including the Ti $3d$ and O $2p$ orbitals (totally, 14 atomic orbitals) as the basis set \cite{kahn, mattheissPR, mattheissPRB}. 
Parameters to be adjusted are the energy difference between the Ti $3d$ and O $2p$ levels, $\epsilon_p - \epsilon_d$, the crystal-field splitting of the O $2p$ orbitals, $\epsilon_{p\sigma} - \epsilon_{p\pi}$, and Slater-Koster parameters ($pd\sigma$), ($pd\pi$), ($pp\sigma$), and ($pp\pi$) \cite{slater}. 
We chose the TB parameters to reproduce the observed band dispersions as well as the indirect and direct optical band gaps of 3.3 eV and 3.8 eV, respectively \cite{vonBenthema} as shown in Fig.~\ref{ARPESwithcalc} (a). 
Best-fit parameters are $\epsilon_p - \epsilon_d$ = $-5.7$ eV, $\epsilon_{d\sigma} - \epsilon_{d\pi}$= $2.2$ eV, $\epsilon_{p\sigma} - \epsilon_{p\pi}$ = $1.5$ eV, ($pd\sigma$) = $-2.3$ eV, ($pp\sigma$) = $0.40$ eV. 
Also, we assumed ($pd\pi$) = $-0.46$($pd\sigma$), ($pp\pi$) = $-0.25$($pp\sigma$) \cite{harrison}. 
Here, the same shift of $\sim 500$ meV between the ARPES peaks and the TB band dispersions are assumed not only at the VBM at the M and R points but also over the entire O $2p$ bands, although the shifts may differ between the bands and $k$ points. 
As shown in Fig.~\ref{ARPESwithcalc} (b), (c), and (d), the calculated band dispersions could successfully reproduce the experimental dispersions of peak positions in the EDCs, except for the overall shift of $\sim 500$ meV between the ARPES data and the TB calculation. 
Here, it should be noted again that the shift of $E_F$ due to electron doping is as small as $\sim 40$ meV as mentioned above, and that according to the previous PES measurement of La$_{1-x}$Sr$_x$TiO$_3$ \cite{Yoshidamaster}, electron doping into STO induces chemical potential shifts as small as $\sim 200$ meV from STO to LaTiO$_3$ (100$\%$ electron doping). 
Therefore, electron doping caused by the oxygen vacancies and/or Nb substitution cannot explain the energy shift as large as $\sim 500$ meV. 

\begin{figure}[!b]
\begin{center}
\includegraphics[width=\linewidth]{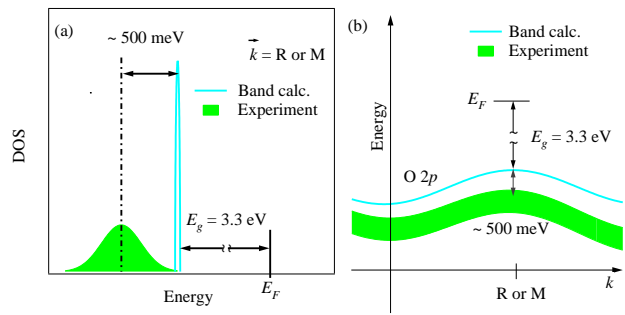}
\caption{(Color online) Schematic illustrations of the strong electron-phonon coupling effects in STO. (a) DOS. (b) Band dispersion. }
\label{shift}
\end{center}
\end{figure}
In order to explain the discrepancies between the ARPES peak dispersion and the TB band structure fitted to the optical band gaps, we suggest that the effect of strong electron-phonon coupling on the spectral line shape is important for the O $2p$ band, too, as schematically shown in Fig.~\ref{shift}. 
We note that the VBM at the M and R points have pure O $2p$ character without admixture of Ti $3d$. 
In the scenario of strong electron-phonon coupling effect on photoemission spectra, strong multiple phonon peaks appear on the higher binding energy side of the weak (almost invisible) zero-phonon line, making the spectral line shape broad and apparently shifted toward higher binding energies by the multiple phonon energy compared with the zero phonon peak (threshold) \cite{elph1D}. 
Therefore, ARPES peaks will be observed at higher binding energies than the optical band gap suggests. 
In the photoemission spectra of sveral TMO's, such effects of electron-phonon coupling have been reported in $3d$-electron photoemission spectra. 
For example, in the case of Ca$_2$CuO$_2$Cl$_2$, the energy shift between the photoemission threshold and the photoemission peak is found to be $\sim 450$ meV \cite{k.shen}, and in the case of La$_{1-x}$Sr$_x$FeO$_3$ $\sim 1$ eV \cite{Wadati-LSFOarpes}. 
When an electron is strongly coupled with optical phonon modes of energy $\omega$ \cite{mahan}, an energy shift of $\Delta$ = $g\omega$ will arise, where $g$ is the electron-phonon coupling constant. 
$g$ is much larger than one because the coupling between the positive charge of the photohole exited in the photoemission process and surrounding ions can be very large. 
It is interesting to note that we have observed effects of similar magnitude for the emission of an O $2p$ electron, which are generally thought to be itinerant weakly correlated. 
The electron-phonon coupling effects of similar magnitudes between the transition-metal $3d$ and O $2p$ electrons mean that a hole in the O $2p$ band created by photoemission has tendency to be localized like a hole in the TM $3d$ bands \cite{Wadati-LSFOarpes}. 

In summary, we have performed detailed ARPES measurements of Nb-doped (lightly electron-doped) STO and have observed the ARPES peak dispersions of the O $2p$ band and the bottom of the Ti 3$d$ band. 
The photoemission threshold at the valence band maxima (the R and M points in the Brillouin zone) were observed at 3.3 eV below $E_{F}$, in agreement with the optical band gap. 
The corresponding photoemission peak, however, was found $\sim 500$ meV below the threshold. 
We interpret the shift of the ARPES peak of $\sim 500$ meV as an effect of strong electron-phonon coupling in analogy with the similar behaviors of $d$-derived photoemission peak in insulating $3d$ TMOs. 

This work was supported by a Giant-in-Aid for Scientific Research (19204037) from JSPS and that for Priority Area ``Invention of Anomalous Quantum Materials" (16076208) from MEXT, Japan. 
The work at KEK-PF was done under the approval of Photon Factory Program Advisory Committee (Proposal No. 2005S2-002).


\begin{thebibliography}{10}

\bibitem{MIT}
M. Imada, A. Fujimori, and Y. Tokura, Rev. Mod. Phys. {\bf 70},  1039  (1998).

\bibitem{k.shen}
K.~M. Shen, F. Ronning, D.~H. Lu, W.~S. Lee, N.~J.~C. Ingle, W. Meevasana, F.
  Baumberger, A.~Damascelli, N.~P. Armitage, L.~L. Miller, Y. Kohsaka, M. Azuma,
  M. Takano, H. Takagi, and Z.~X. Shen, Phys. Rev. Lett. {\bf 93},  267002
  (2004).

\bibitem{Wadati-LSFOarpes}
H. Wadati, A. Chikamatsu, M. Takizawa, R. Hashimoto, H. Kumigashira, T.
  Yoshida, T. Mizokawa, A. Fujimori, M. Oshima, M. Lippmaa, M. Kawasaki, and H.
  Koinuma, Phys. Rev. B {\bf 74},  115114  (2006).

\bibitem{okazaki}
K. Okazaki, H. Wadati, A. Fujimori, M. Onoda, Y. Muraoka, and Z. Hiroi, Phys.
  Rev. B {\bf 69},  165104  (2004).

\bibitem{mishchenko}
A.~S. Mishchenko and N. Nagaosa, Phys. Rev. Lett. {\bf 93},  036402  (2004).

\bibitem{Rosch}
O. R\"osch, O. Gunnarsson, X.~J. Zhou, T. Yoshida, T. Sasagawa, A. Fujimori, Z.
  Hussain, Z.-X. Shen, and S. Uchida, Phys. Rev. Lett. {\bf 95},  227002
  (2005).

\bibitem{Goldschmidt}
D. Goldschmidt and H.~L. Tuller, Phys. Rev. B {\bf 35},  4360  (1987).

\bibitem{vonBenthema}
K. van Benthem, C. Els\"asser, and R.~H. French, J. Appl. Phys. {\bf 90},  6156
   (2001).

\bibitem{STO-SC1}
J.~F. Schooley, W.~R. Hosler, and M.~L. Cohen, Phys. Rev. Lett. {\bf 12},  474
  (1964).

\bibitem{STO-SC2}
C.~S. Koonce, M.~L. Cohen, J.~F. Schooley, W.~R. Hosler, and E.~R. Pfeiffer,
  Phys. Rev. {\bf 163},  380  (1967).

\bibitem{gervais}
F. Gervais, J.~L. Servoin, A.~Baratoff, J.~G. Bednorz, and G. Binnig, Phys.
  Rev. B {\bf 47},  8187  (1993).

\bibitem{eagles}
D.~M. Eagles, M. Georgiev, and P.~C. Petrova, Phys. Rev. B {\bf 54},  22
  (1996).

\bibitem{ang}
ChenAng, ZhiYu, ZhiJing, P. Lunkenheimer, and A. Loidl, Phys. Rev. B {\bf 61},
  3922  (2000).

\bibitem{Bi}
C.~Z. Bi, J.~Y. Ma, J. Yan, X. Fang, B.~R. Zhao, D.~Z. Yao, and X.~G. Qiu, J.
  Phys.: Condens. Mat. {\bf 18},  2553  (2006).

\bibitem{mattheissPRB}
L.~F. Mattheiss, Phys. Rev. B {\bf 6},  4718  (1972).

\bibitem{mattheissPR}
L.~F. Mattheiss, Phys. Rev. {\bf 181},  987  (1969).

\bibitem{kahn}
A.~H. Kahn and A.~J. Leyendecker, Phys. Rev. {\bf 135},  A1321  (1964).

\bibitem{haruyama}
Y. Haruyama, S. Kodaira, Y. Aiura, H. Bando, Y. Nishihara, T. Maruyama, Y.
  Sakisaka, and H. Kato, Phys. Rev. B {\bf 53},  8032  (1996).

\bibitem{Aiura}
Y. Aiura, I. Hase, H. Bando, T. Yasue, T. Saitoh, and D.~S. Dessau, Surf. Sci.
  {\bf 515},  61  (2002).

\bibitem{kawasaki}
M. Kawasaki, K. Takahashi, T. Maeda, R. Tsuchiya, M. Shinohara, O. Ishiyama, T.
  Yonezawa, M. Yoshimoto, and H. Koinuma, Science {\bf 266},  1540  (1994).

\bibitem{horiba}
K. Horiba, H. Ohguchi, H. Kumigashira, M. Oshima, K. Ono, N. Nakagawa, M.
  Lippmaa, M. Kawasaki, and H. Koinuma, Rev. Sci. Instr. {\bf 74},  3406
  (2003).

\bibitem{wadatitbc}
H. Wadati, T. Yoshida, A. Chikamatsu, H. Kumigashira, M. Oshima, H. Eisaki,
  Z.~X. Shen, T. Mizokawa, and A. Fujimori, Phase Transitions {\bf 79},  617
  (2006).

\bibitem{slater}
J.~C. Slater and G.~F. Koster, Phys. Rev. {\bf 94},  1498  (1954).

\bibitem{harrison}
W.~A. Harrison, {\em Electronic structure and the properties of solids} (Dover,
  New York, 1989).

\bibitem{Yoshidamaster}
T. Yoshida, Master's thesis, University of Tokyo, 1999.

\bibitem{elph1D}
L. Perfetti, H. Berger, A. Reginelli, L. Degiorgi, H. H\"ochst, J. Voit, G.
  Margaritondo, and M. Grioni, Phys. Rev. Lett. {\bf 87},  216404  (2001).

\bibitem{mahan}
G.~D. Mahan, {\em Many-Particle Physics} (Plenum, New York, 1981).

\end{thebibliography}
\end{document}